\documentclass[conference]{IEEEtran}
\IEEEoverridecommandlockouts
\usepackage{subfigure}
\usepackage{cite}
\usepackage{amsmath,amssymb,amsfonts}
\usepackage{algorithmic}
\usepackage{algorithm}
\usepackage{graphicx}
\usepackage{textcomp}
\usepackage{xcolor}
\usepackage{url}
\usepackage{bm}
\begin{document}

\title{Multi-hop Parallel Image Semantic Communication for Distortion Accumulation Mitigation}
\author{Bingyan Xie, Jihong Park, Yongpeng Wu, Wenjun Zhang, Tony Q.S. Quek

\thanks{This work has been supported in part by the SNS JU project 6G-GOALS under the EU’s Horizon program Grant Agreement No. 101139232.}
\thanks{Bingyan Xie, Yongpeng Wu, and Wenjun Zhang are with the Department of Electronic Engineering, Shanghai Jiao Tong University, Shanghai 200240, China (e-mail:bingyanxie, yongpeng.wu, fengbiqian, zhangwenjun@sjtu.edu.cn).}% <-this % stops a space
\thanks{Jihong Park and Tony Q.S. Quek are with the ISTD Pillar, Singapore University of Technology of Design, 8 Somapah Rd, Singapore 487372 (e-mail:jihong\_park, tonyquek@sutd.edu.sg)}
}
\maketitle
\begin{abstract}
Existing semantic communication schemes primarily focus on single-hop scenarios, overlooking the challenges of multi-hop wireless image transmission. As semantic communication is inherently lossy, distortion accumulates over multiple hops, leading to significant performance degradation. To address this, we propose the multi-hop parallel image semantic communication (MHPSC) framework, which introduces a parallel residual compensation link at each hop against distortion accumulation. To minimize the associated transmission bandwidth overhead, a coarse-to-fine residual compression scheme is designed. A deep learning-based residual compressor first condenses the residuals, followed by the adaptive arithmetic coding (AAC) for further compression. A residual distribution estimation module predicts the prior distribution for the AAC to achieve fine compression performances. This approach ensures robust multi-hop image transmission with only a minor increase in transmission bandwidth. Experimental results confirm that MHPSC outperforms both existing semantic communication and traditional separated coding schemes.
\end{abstract}


\section{Introduction}
Semantic communication, an emerging paradigm for efficiently transmitting multi-media data, is widely envisioned as a promising technology of 6G networks. Unlike traditional separated source-channel coding (SSCC), this approach represents a significant advancement by deeply integrating artificial intelligence (AI) into communication system designs. Leveraging deep learning (DL) networks, semantic communication focuses on extracting and transmitting the underlying semantic meaning of data, rather than process raw pixels in images or videos. This shift in focus substantially reduces communication overhead. Consequently, it is increasingly applied in diverse fields such as the Internet of things, smart cities, and extended reality.

The prevailing architecture for semantic communication frameworks is derived from the idea of joint source-channel coding (JSCC), which has been extensively applied to wireless image transmission tasks. For example, Xu et al. \cite{ADJSCC} have embedded an attention-based SNR-adaptive module into the JSCC backbone to maintain performance across diverse signal-to-noise ratios (SNRs). Xie et al. \cite{LCFSC} have improved image transmission robustness against MIMO interference by fusing channel state information into the semantic encoder. Diverging from these single-link approaches, Xu et al. \cite{parasc} have proposed a hybrid system that incorporates a parallel SSCC link, facilitating joint decoding that leverages both semantic and conventional data streams.

While existing semantic communication schemes successfully reduce communication costs and optimize image reconstruction performance, they are predominantly designed for single-hop scenarios in a point-to-point manner. This overlooks the more prevalent multi-hop networks in practice. Directly deploying these single-hop schemes in multi-hop settings leads to severe distortion accumulation, as semantic transmission is inherently lossy. An et al. \cite{relay} have proposed a relay-based image transmission system with a single relay node. Although this two-hop scheme employs hyperprior entropy recompression for the first hop, it neglects to address the performance degradation caused by accumulated distortion. To address this, Zhang et al. \cite{mhdeepsc} have proposed a multi-hop framework featuring a recursive training method to mitigate this distortion. Their approach, which makes each hop aware of previous ones through joint training, provides positive performance gain to some extent. However, the reconstructed image quality degrades significantly as the hop count rises. This is because the semantic codecs in \cite{mhdeepsc} utilize identical structures and weights across all nodes. Consequently, a mere training strategy improvement fails to address the fundamental limitations of the underlying single-hop architecture against distortion brought by multiple hops.

Based on the above analysis, there is an urgent requirement to design a robust wireless image semantic transmission framework in multi-hop scenarios. To this means, we propose the multi-hop parallel semantic communication (MHPSC) framework, to efficiently mitigate the problem of distortion accumulation. In addition to the standardized single-hop link, MHPSC introduces an extra parallel residual compensation link. Unlike the digital parallel link in \cite{hsc} utilized to improve controllable fidelity for generative semantic communication with a single hop, proposed compensation link counteracts the distortion from the previous hop by injecting calculated residuals into the received image at each node. To minimize the communication overhead, the designed compensation link introduces a coarse-to-fine residual compression scheme, adapting both DL-based residual compressor and adaptive arithmetic coding (AAC) to compress transmitted residuals to a great extent. Consequently, MHPSC achieves significant robustness and performance gains in multi-hop scenarios with only a minor addition to the single-link transmission cost. The contributions of this paper are as follows

\begin{figure*}[htbp]
	\centering
	\includegraphics[width=5.2in]{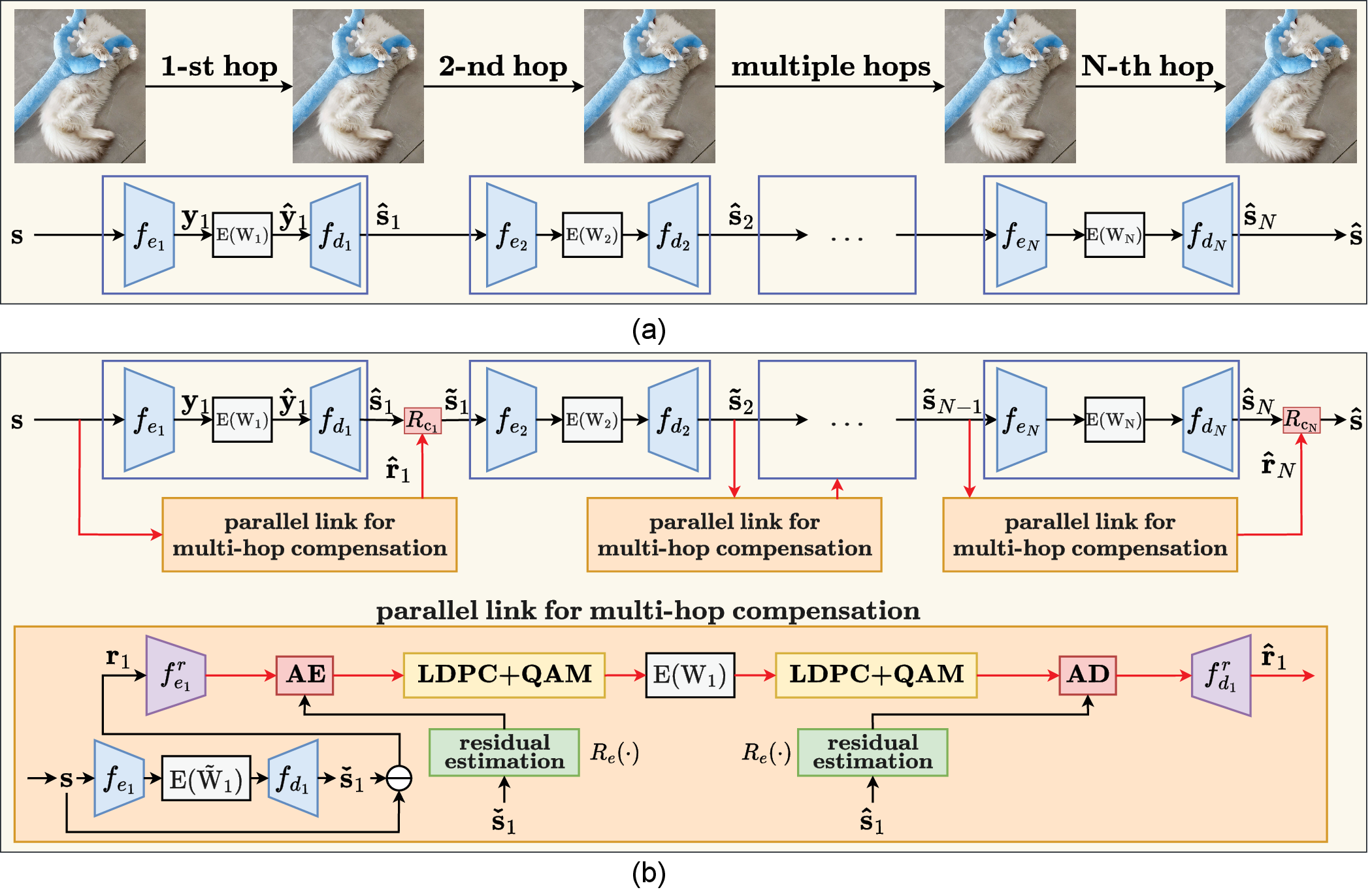}
	\caption{The system model for the multi-hop wireless image transmission scenario. (a) The common multi-hop wireless image transmission link. (b) The proposed MHPSC with parallel link for multi-hop compensation. The red lines are the residual transmission process.}
	\label{fig_1}
\end{figure*}
\begin{enumerate}
	\item{}	
	To alleviate the distortion accumulation in multi-hop transmission, we propose a multi-hop parallel image semantic communication framework which incorporates a unique residual compensation link to mitigate the information loss introduced by previous hops.
	\item{}	
	To minimize the transmission cost of the residual compensation link, we propose a coarse-to-fine residual compression scheme. This scheme first uses a DL-based compressor for primary compression. A residual estimation module then predicts the residual distribution, which guides the AAC module to achieve further compression. This efficient two-stage process ensures robust performance against distortion accumulation with only a minor increase in transmission cost.
	\item{}
	To demonstrate the effectiveness of proposed MHPSC, numerical experiments are conducted. DL-based semantic communication schemes \cite{ADJSCC, mhdeepsc, WITT} along with traditional SSCC schemes are compared to certificate the superior performance of MHPSC in multi-hop scenarios.
\end{enumerate}

Notations: $\mathbb{R}$ and $\mathbb{C}$ refer to the real and complex number sets, respectively. $\mathcal{CN}\left (\mu, \sigma^2 \right)$ denotes a complex Gaussian distribution with mean $\mu$ and variance $\sigma^2$. $\mathrm{diag}(\cdot)$ refer to the diagonalization operations between a vector and its corresponding diagonal matrix. $|\cdot|$ refers to computing the modulus of a complex number. $\cdot$ refers to the element-wise multiplication. $\mathbf{I}$ denotes the unit matrix. The operator $\left(\cdot\right)^{T}$ denotes the matrix transpose.

\section{System Model and Proposed Framework}
In this section, we first describe the common multi-hop wireless image transmission system model and then construct proposed MHPSC with the parallel transmission link.

\subsection{Common Multi-hop Wireless Image Transmission Model}
We consider a multi-hop wireless image transmission scenario, where images are carried along multiple nodes through wireless links. For an arbitrary hop $n$ with $n=1,\cdots, N$, the encoder, $f_{e_n}(\cdot): \mathbb{R}^{H\times W\times 3}\longrightarrow \mathbb{R}^{L}$, encodes the received image, $\mathbf{s}_n$, into a codeword sequence, $\mathbf{y}_n\in \mathbb{R}^{L}$, with code length $L$. After that, the codewords pass through the Rayleigh fading channel along with minimum mean square error (MMSE) equalization, which is formulated as
\begin{align}
	\hat{\mathbf{y}}_n=\mathbf{H}_{s_n}\cdot\mathbf{y}_n+\mathbf{H}_{n_n}\cdot\mathbf{n}_n, 
\end{align}
where $\hat{\mathbf{y}}_n$ is the received codewords with the $n$-th hop, $\mathbf{H}_{s_n}=\mathrm{diag}(|\mathbf{h}_{d_n}|^2(|\mathbf{h}_{d_n}|^2+\sigma^{2}\mathbf{I})^{-1})\in\mathbb{R}^{L}$ and $\mathbf{H}_{n_n}=\mathrm{diag}(\mathbf{h}_{d_n}^T(|\mathbf{h}_{d_n}|^2+\sigma^{2}\mathbf{I})^{-1})\in\mathbb{C}^{L}$ refer to the channel equalization parameters, $\mathbf{h}_{d_n} = \mathrm{diag}(\mathbf{h}_n)\in \mathbb{C}^{L\times L}$, $\mathbf{h}_n\in \mathbb{C}^{L}$ denotes the Rayleigh channel fading index following the distribution of $\mathcal{CN}(0,1)$, $\mathbf{n}_n\in \mathbb{C}^{L}$ is the complex Gaussian channel noise vector whose component has zero mean and covariance $\sigma^{2}_n$.

With $\hat{\mathbf{y}}_n$, the decoder, $f_{d_n}(\cdot): \mathbb{R}^{L}\longrightarrow \mathbb{R}^{H\times W\times 3}$, converts $\hat{\mathbf{y}}_n$ into the reconstructed image $\hat{\mathbf{s}}_n$. The transmission process for an arbitrary hop $n$ can be formulated as
\begin{align}
	\mathbf{s}_n \xrightarrow{f_{e_n}(\cdot)} \mathbf{y}_n \xrightarrow{E\left(W_n(\cdot)\right)} \mathbf{\hat{y}}_n \xrightarrow{f_{d_n}(\cdot)} \mathbf{\hat{s}}_n,
\end{align}
where $W_n(\cdot)$ implies wireless channels of the $n$-th hop and $E(\cdot)$ refers to the MMSE channel equalization.

Then we extend the single-hop to the multi-hop semantic transmission. As shown in Fig. \ref{fig_1}(a), the multi-hop transmission starts from a raw image $\mathbf{s}\in\mathbb{R}^{H\times W\times3}$, where $\mathbf{s}_1=\mathbf{s}$. For the $n$-th hop, the transmitted image $\mathbf{s}_n$ is the received image from the ($n-1$)-th hop as $\mathbf{s}_n=\mathbf{\hat{s}}_{n-1}$. After delivering with $N$ successive hops, the final received image, $\hat{\mathbf{s}}=\hat{\mathbf{s}}_N$, can be obtained by node $N$. Note that the network model of each hop shares the same structure and parameter weights as assumed in \cite{mhdeepsc}. The whole multi-hop wireless image transmission process can be formulated as
\begin{align}
	\mathbf{s}_1 \xrightarrow{\textbf{1-st hop}} \mathbf{\hat{s}}_1 \xrightarrow{\cdots} \mathbf{\hat{s}}_n\xrightarrow{\cdots} \mathbf{\hat{s}}_{N-1} \xrightarrow{\textbf{N-th hop}} \mathbf{\hat{s}}_N.
\end{align}

\subsection{Proposed Multi-hop Parallel Image Semantic Communication Framework}
In common multi-hop wireless image transmission model, images are delivered hop-by-hop. This allows the distortion introduced at each hop to amplify with increasing hop count, leading to severe performance degradation for the end node. To combat distortion accumulation, we construct a multi-hop parallel image semantic communication framework based on the aforementioned common multi-hop transmission model.

Unlike the single-link structure for multi-hop transmission, proposed MHPSC employs two parallel transmission links. One is the common multi-hop link as presented in Fig. \ref{fig_1}(a). The other is the unique residual compensation link which is utilized to combat the accumulated distortion brought by the previous hops. As shown in Fig. \ref{fig_1}(b), for the $n$-th hop transmission, the semantic coder $f_{e_n}(\cdot)$ and $f_{d_n}(\cdot)$ preprocess the original image $\mathbf{s}$ into $\mathbf{\check{s}}_n$, where $\tilde{W}_n(\cdot)$ refers to the emulated channel with the same channel parameters, $\mathbf{h}_n$ and SNR. Then the residuals, $\mathbf{r}_n\in\mathbb{R}^{H\times W\times 3}$, are computed as $\mathbf{r}_n=\mathbf{s}_n-\mathbf{\check{s}}_n$. Before transmitting residuals, the residual compressor $f_{e_n}^r(\cdot)$ encodes residuals into $\mathbf{\tilde{r}}_n$. With compressed $\mathbf{\tilde{r}}_n$, we design the AAC to further compress the transmitted residuals. To exploit the characteristic of AAC for data compression, the DL-based residual estimation module is employed to learn the residual distribution as $p$. With both $\mathbf{\tilde{r}}_n$ and corresponding distribution, we are able to efficiently transmit residuals through such SSCC scheme. 'AE' and 'AD' refers to the arithmetic coder pair for compress and recover the residuals. 'LDPC+QAM' refers to the low density parity check (LDPC) channel coding and quadrature amplitude modulation (QAM). In this way, residuals are compressed, encoded and mapped to a series of discrete constellation points as $\mathbf{r}_n^p\in\mathbb{C}^{\mathrm{L_r}}$. The residual transmission is formulated as
\begin{align}
	\mathbf{{\hat{r}}}_n^p=\mathbf{H}_{s_n^r}\cdot\mathbf{r}_n^p+\mathbf{H}_{n_n^r}\cdot\mathbf{n}_n^r, 
\end{align}
where $\mathbf{{\hat{r}}}_n^p$ refers to the received compressed residuals, the definitions of $\mathbf{H}_{s_n^r}=\mathrm{diag}(|\mathbf{h}_{d_n^r}|^2(|\mathbf{h}_{d_n^r}|^2+\sigma^{2}\mathbf{I})^{-1})\in\mathbb{R}^{L^r}$, $\mathbf{H}_{n_n^r}=\mathrm{diag}(\mathbf{h}_{d_n^r}^T(|\mathbf{h}_{d_n^r}|^2+\sigma^{2}\mathbf{I})^{-1})\in\mathbb{C}^{L^r}$, $\mathbf{h}_{d_n^r} = \mathrm{diag}(\mathbf{h}_n^r)\in \mathbb{C}^{L^r\times L^r}$ and $\mathbf{h}_n^r\in \mathbb{C}^{L^r}$ are similar to Eq. (1).

Then, the QAM demodulator, SSCC decoder, and the residual decompressor $f_{d_n}^r(\cdot)$ transform received residuals into reconstructed residuals $\mathbf{{\hat{r}}}_n\in\mathbb{R}^{H\times W\times 3}$. Finally, the reconstructed image is compensated $\mathbf{\tilde{s}}_n$ as
\begin{align}
	\mathbf{{\tilde{s}}}_n=R_{\mathrm{c}_n}(\mathbf{\hat{s}}_n,\mathbf{\hat{r}}_n).
\end{align}
where $R_{\mathrm{c}_n}$ refers to the residual combination network.

By incorporating such parallel stream for multi-hop compensation, the proposed framework significantly alleviates distortion accumulation.

\subsection{Residual Estimation}
Since AAC inputs source distribution as prior knowledge to achieve superior lossless compression performance, accurate residual distribution is required for both the encoder and decoder end of the residual compensation link. In this way, the residual estimation module $R_e(\cdot)$ is introduced to provide distribution for the arithmetic coder. Given $\mathbf{\check{s}}_n$, $R_e(\cdot)$ predicts the probability mass function of the residual $\mathbf{\tilde{r}}_n$ as
\begin{align}
	p(\mathbf{\tilde{r}}_n|\mathbf{\check{s}}_n)=R_e(\mathbf{\check{s}}_n).
\end{align}

Accord to \cite{rc}, a discrete mixture of logistic distributions is employed to model $\mathbf{\tilde{r}}_n$. Let $c=1,2,3$ denotes the red green blue (RGB) channels and $u$,$v$ the spatial location. Eq. (6) can be further defined as
\begin{align}
	p(\mathbf{\tilde{r}}_n|\mathbf{\check{s}}_n)=\prod_{u,v}p(\mathbf{\tilde{r}}_n^{1uv},\mathbf{\tilde{r}}_n^{2uv},\mathbf{\tilde{r}}_n^{3uv}|\mathbf{\check{s}}_n).
\end{align}

We use a weak autoregression over the three RGB channels to define the joint distribution over channels via logistic mixtures $p_m$ as
\begin{equation}
	\begin{aligned}
		p(\mathbf{\tilde{r}}_n^{1uv},\mathbf{\tilde{r}}_n^{2uv},\mathbf{\tilde{r}}_n^{3uv}|\mathbf{\check{s}}_n)=&p_m(\mathbf{\tilde{r}}_n^{1uv}|\mathbf{\check{s}}_n)\cdot p_m(\mathbf{\tilde{r}}_n^{2uv}|\mathbf{\check{s}}_n,\mathbf{\tilde{r}}_n^{1uv})\cdot\\& p_m(\mathbf{\tilde{r}}_n^{3uv}|\mathbf{\check{s}}_n,\mathbf{\tilde{r}}_n^{1uv},\mathbf{\tilde{r}}_n^{2uv}).
	\end{aligned}
\end{equation}

Then, a mixture of $K = 5$ $(k=1,2,3,4,5)$ logistic distributions $p_L$ is used to formulate $p_m$. The distributions are defined by the outputs of $R_e(\cdot)$, which yields mixture weights $\pi^{k,cuv}_n$, means $\mu^{k,cuv}_n$, variances $\sigma^{k,cuv}_n$, and mixture coefficients $\lambda^{k,cuv}_n$. The autoregression over RGB channels is only used to update the means using a linear combination of $\mu_n$ and the target $\mathbf{\tilde{r}}_n$ of previous channels, scaled by the coefficients $\lambda_n$. We thereby obtain $\tilde{\mu}_n$ as
\begin{align}
	\tilde{\mu}_n^{k,uv}= \begin{cases}
		\mu_n^{k,1uv}, & c=1\\
		\mu_n^{k,2uv}+\lambda_n^{k,1uv}\mathbf{\tilde{r}}_n^{k,1uv},  & c=2\\
		\mu_n^{k,3uv}+\lambda_n^{k,2uv}\mathbf{\tilde{r}}_n^{k,1uv}+\lambda_n^{k,3 uv}\mathbf{\tilde{r}}_n^{k,2uv},  & c=3
	\end{cases}
\end{align}

As the logistic mixtures, $p_m$ can be formulated as
\begin{align}
	p_m(\mathbf{\tilde{r}}_n^{cuv}|\mathbf{\check{s}}_n,\mathbf{\tilde{r}}_n^\mathrm{prev})=\sum_{k=1}^{K}\pi^{k,cuv}_n p_L(\mathbf{\tilde{r}}_n^{cuv}|\tilde{\mu}^{k,cuv}_n,\sigma^{k,cuv}_n),
\end{align}
where $\mathbf{\tilde{r}}_n^\mathrm{prev}$ denotes the channels with index smaller than $c$.

Among this, $p_L$ is the logistic distribution presented as
\begin{align}
	p_L(\mathbf{\tilde{r}}_n|\tilde{\mu}^{k,cuv}_n,\sigma^{k,cuv}_n)=\frac{e^{-(\mathbf{\tilde{r}}_n-\tilde{\mu}^{k,cuv}_n)/\sigma^{k,cuv}_n}}{\sigma^{k,cuv}_n(1+e^{-(\mathbf{\tilde{r}}_n-\tilde{\mu}^{k,cuv}_n)/\sigma^{k,cuv}_n})^2}.
\end{align}

We evaluate $p_L$ at discrete $\mathbf{\tilde{r}}_n$, via its cumulative distribution function (CDF) as
\begin{align}
	p_L(\mathbf{\tilde{r}}_n)=\mathrm{CDF}(\mathbf{\tilde{r}}_n+\frac{1}{2})-\mathrm{CDF}(\mathbf{\tilde{r}}_n-\frac{1}{2}).
\end{align}

\section{Deployment Detail}
In this section, we present the network structure and training strategy of the MHPSC framework.

\subsection{Network Structure}
\begin{figure}[htbp]
	\centering
	\includegraphics[width=2.8in]{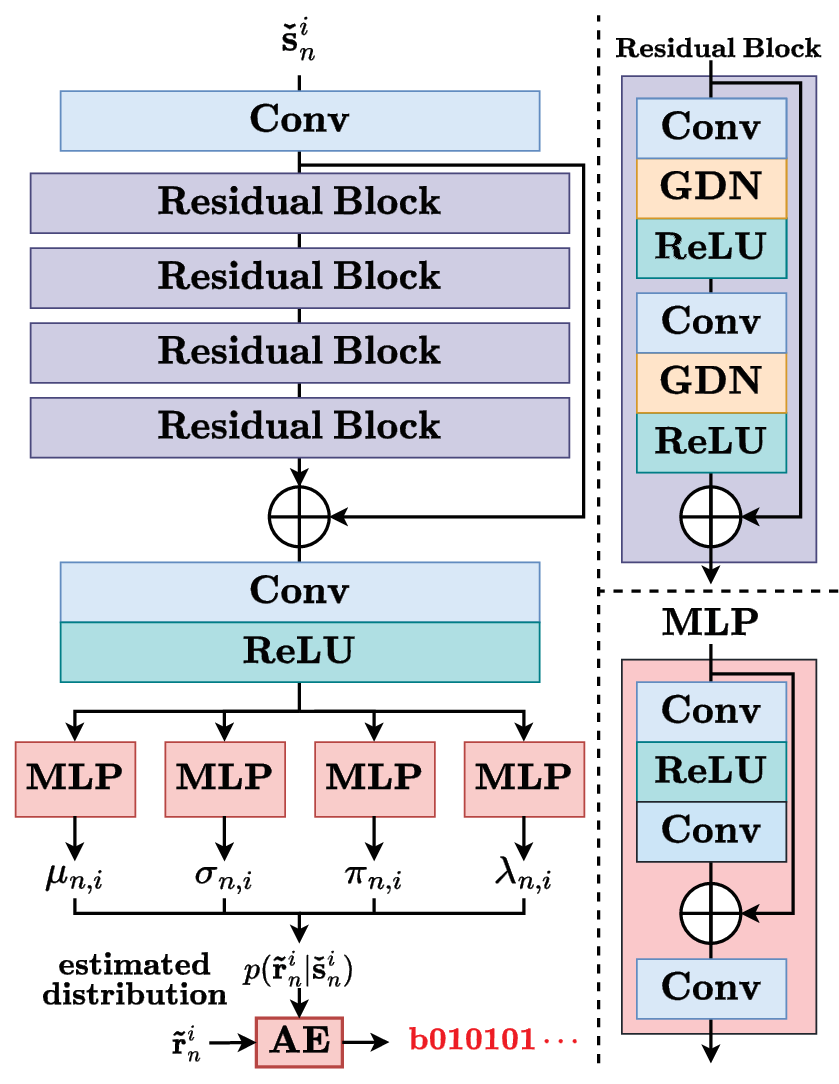}
	\caption{The residual estimation module for predicting distributions.}
	\label{fig_2}
\end{figure}

In accordance to Fig. \ref{fig_1}(b), the main network backbone is composed of \cite{WITT}. The residual compressor is the same as the residual encoder and decoder in \cite{WVSC}. The residual combination network follows the same U-net structure as \cite{cvst}. For the residual estimation module, it is shown in Fig. \ref{fig_2}. It is composed of a series of residual blocks which mainly contain the convolution layer, generalized divisive normalization layer (GDN), and ReLU layer. After the concatenation through residual blocks, four multi-layer perceptron (MLP) layers finally predict all the required $\mu_{n,i}$, $\sigma_{n,i}$, $\pi_{n,i}$, and $\lambda_{n,i}$ for the discrete mixture of logistic distributions, where $i$ refers to the $i$-th raw image.

\subsection{Training Loss and Strategy}
For the model training, we combine the idea of multi-hop training with the single-hop training with three stages. For the first stage, we train solely the network backbone $\mathbf{f}_{e_n}(\cdot)$ and $\mathbf{f}_{d_n}(\cdot)$ with multiple hops enhanced by the recursive training strategy \cite{mhdeepsc} as
\begin{align}
	L_\mathrm{1}=\frac{1}{N\times I}\sum_{i=1}^{I}\sum_{n=1}^{N}(\gamma^{N-n}D\left(\mathbf{s}_n^i,\mathbf{\hat{s}}_n^i\right)),
\end{align}
where $\gamma$ refers to the scaling factor for adjusting the weight of each hop. $I$ is the total number of images. $D(\cdot,\cdot)$ refers to the loss term which is set as mean square error (MSE) in default.

For the second stage, the network backbone parameters are fixed and the residual compressor $\mathbf{f}_{e_n}^r(\cdot)$ and decompressor $\mathbf{f}_{d_n}^r(\cdot)$ are added into the training process. The training loss is the same as Eq. (13).

For the third stage, the residual distribution estimation module $R_e(\cdot)$ is finally added into the network training process to provide accurate residual distribution to the AAC. The training loss is given as
\begin{align}
	L_\mathrm{RC}=-\frac{1}{I}\sum_{i=1}^{I}\log p(\mathbf{\tilde{r}}_n^i|\mathbf{\check{s}}_n^i).
\end{align}

Since the distribution estimation only refers to reduce the channel bandwidth ratio (CBR) from transmitting arithmetic coding bitstreams, we thus employ the single-hop training to simplify the training process. The whole training algorithm is summarized in Alg. 1.

\begin{algorithm}[htbp]
	\caption{Training algorithm for proposed MHPSC}\label{alg:alg1}
	\begin{algorithmic}
		\STATE 
		$\textbf{Input:}$ Raw image $\mathbf{s}_n$, Number of hops N		
		
		$\textbf{Output:}$ Well trained $f_{e_n}(\cdot)$, $f_{d_n}(\cdot)$, $f_{e_n}^r(\cdot)$, $f_{d_n}^r(\cdot)$, $R_e(\cdot)$ 
		
		\STATE \textbf{Training Stage1:}
		\STATE1. for each hop $n = 1,...,N$ do
		\STATE2. \hspace{0.5cm} $\mathbf{s}_n \xrightarrow{f_{e_n}(\cdot)} \mathbf{y}_n \xrightarrow{E\left(W_n(\cdot)\right)} \mathbf{\hat{y}}_n \xrightarrow{f_{d_n}(\cdot)} \mathbf{\hat{s}}_n$
		\STATE3. \hspace{0.5cm} Compute the loss $L_1$ by Eq. (13)
		\STATE4. \hspace{0.5cm} Update the network weights of $f_{e_n}(\cdot)$, $f_{d_n}(\cdot)$

		\STATE \textbf{Training Stage2:}
		\STATE5. for each hop $n = 1,...,N$ do
		\STATE6. \hspace{0.5cm} $\mathbf{s}_n \xrightarrow{f_{e_n}(\cdot)} \mathbf{y}_n \xrightarrow{E\left(\tilde{W}_n(\cdot)\right)} \mathbf{\hat{y}}_n \xrightarrow{f_{d_n}(\cdot)} \mathbf{\check{s}}_n$
		\STATE7. \hspace{0.5cm} $\mathbf{r}_n=\mathbf{s}_n-\mathbf{\check{s}}_n$
		\STATE8. \hspace{0.5cm} $\mathbf{r}_n \xrightarrow{f_{e_n}^r(\cdot)} \mathbf{\tilde{r}}_n \xrightarrow{f_{d_n}^r(\cdot)} \mathbf{\hat{r}}_n$
		\STATE9. \hspace{0.5cm} $\mathbf{{\tilde{s}}}_n=R_\mathrm{c}(\mathbf{\hat{s}}_n,\mathbf{\hat{r}}_n)$
		\STATE10. \hspace{0.3cm} Compute the loss $L_1$ by Eq. (13)
		\STATE11. \hspace{0.3cm} Update the network weights of $f_{e_n}^r(\cdot)$, $f_{d_n}^r(\cdot)$

		\STATE \textbf{Training Stage3:}
		\STATE12. \hspace{0.3cm} $p_n=R_e(\mathbf{\check{s}}_n)$
		\STATE13. \hspace{0.3cm} Compute the loss $L_{\mathrm{RC}}$ by Eq. (14)
		\STATE14. \hspace{0.3cm} Update the network weights of $R_e(\cdot)$
		
		\STATE return the well trained MHPSC		
	\end{algorithmic}
	\label{alg1}
\end{algorithm}

\begin{figure*}[htbp]
	\centering  %图片全局居中
	\subfigure[PNSR for the reconstructed images.]{
		\includegraphics[width=0.30\linewidth]{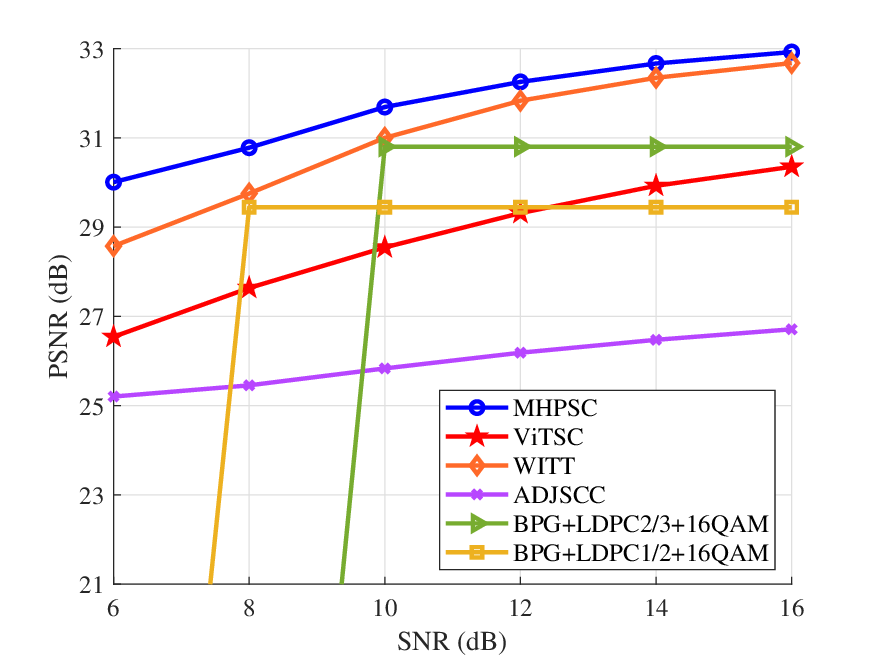}}
	\subfigure[MS-SSIM for the reconstructed images.]{
		\includegraphics[width=0.30\linewidth]{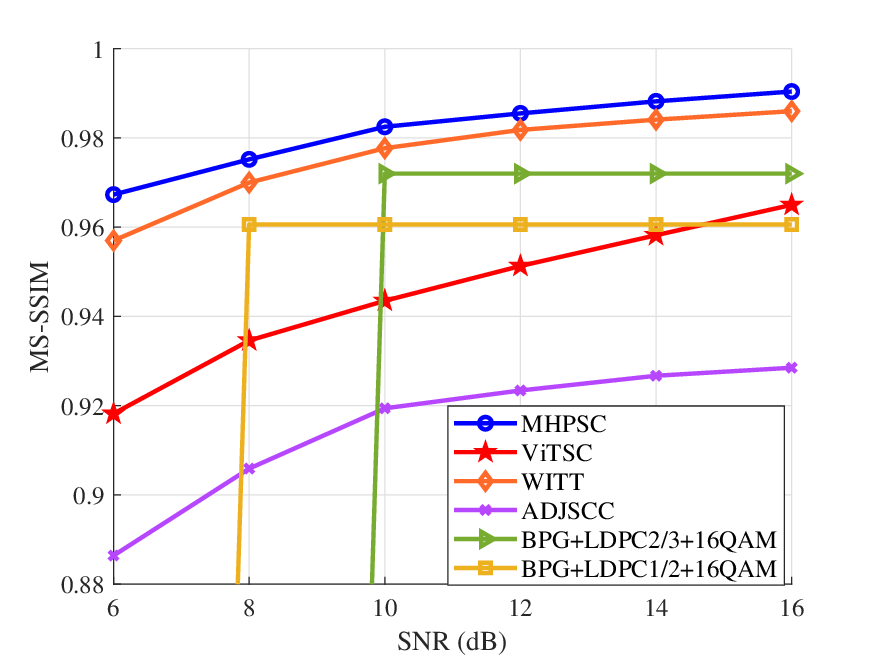}}
	\subfigure[LPIPS for the reconstructed images.]{
		\includegraphics[width=0.30\linewidth]{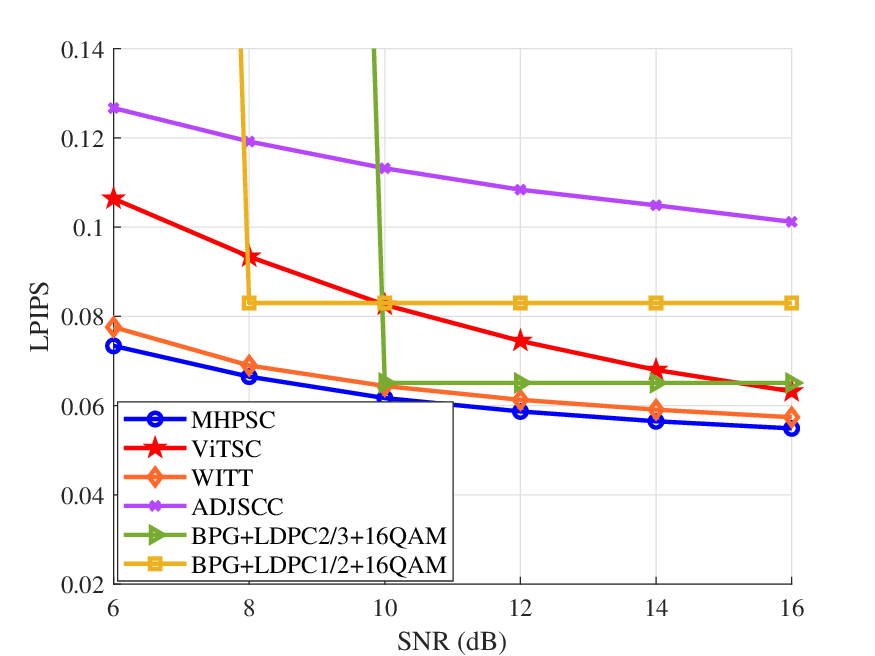}}
	\caption{Quality of the reconstructed images versus the SNRs under Rayleigh fading channels (CBR = 0.077, N = 20).}
	\label{fig_3}
\end{figure*}
\begin{figure*}[htbp]
	\centering  %图片全局居中
	\subfigure[PNSR for the reconstructed images.]{
		\includegraphics[width=0.30\linewidth]{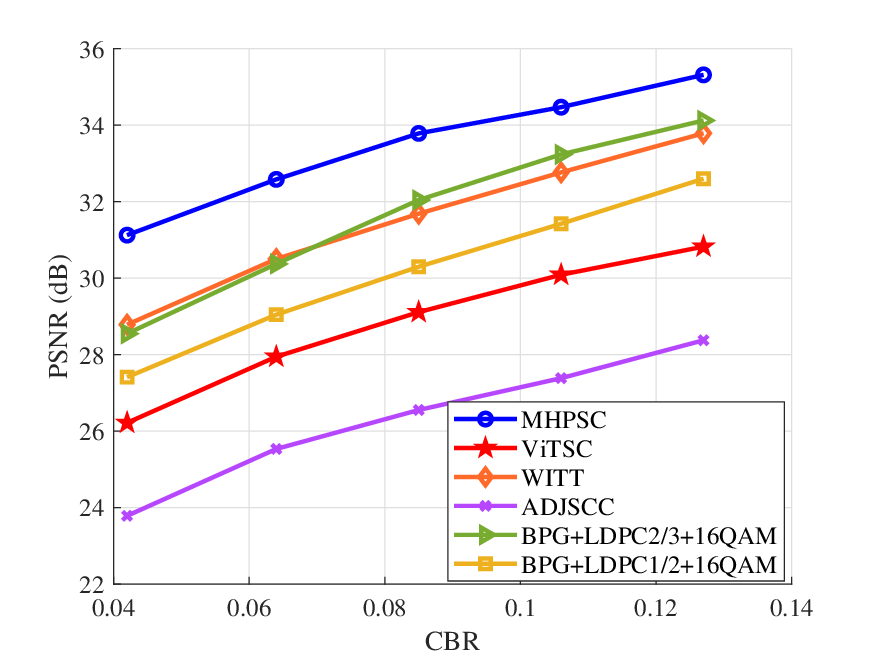}}
	\subfigure[MS-SSIM for the reconstructed images.]{
		\includegraphics[width=0.30\linewidth]{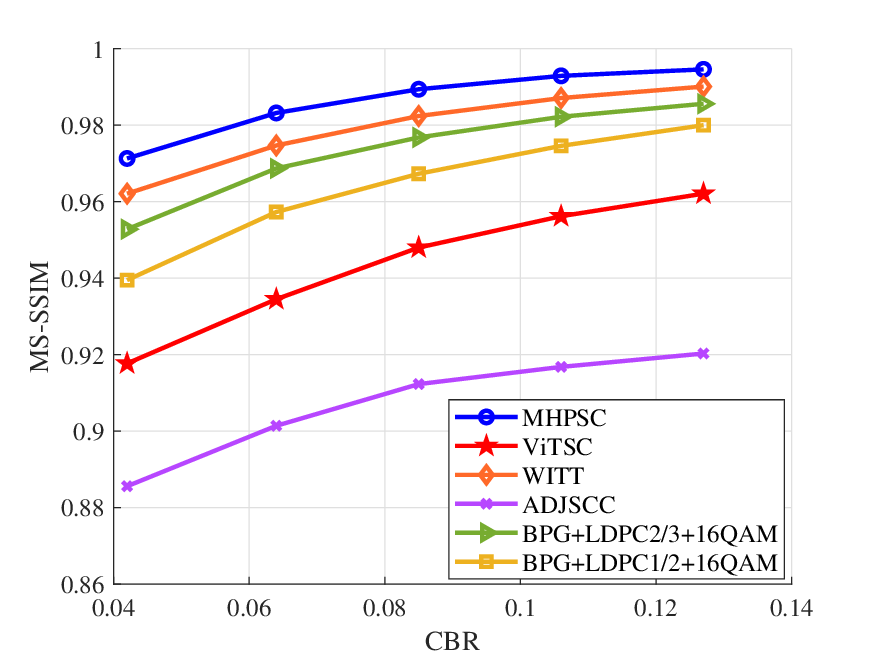}}
	\subfigure[LPIPS for the reconstructed images.]{
		\includegraphics[width=0.30\linewidth]{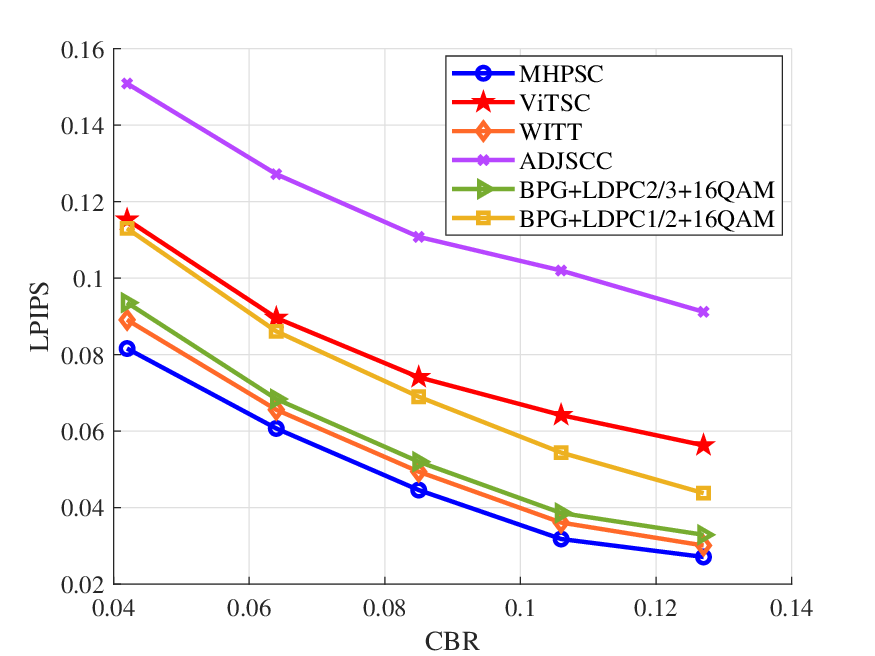}}
	\caption{Quality of the reconstructed images versus the CBRs under Rayleigh fading channels (SNR = 10 dB, N = 20).}
	\label{fig_4}
\end{figure*}

\section{Numerical Results}
In this section, we present numerical results to evaluate the effectiveness of proposed MHPSC for wireless image transmission.

\subsection{Experimental Setups}
\subsubsection{Datasets}

For the wireless image semantic transmission, we quantify the performances of proposed MHPSC versus other benchmarks over the UDIS-D \cite{UDIS-D} dataset, which contains over 10000 real-world images. During model training, images are randomly cropped to 128$\times$128$\times$3.

\subsubsection{Model Deployment Details}
The network deployment of MHPSC utilizes the Swin-Transformer \cite{swin} backbone as the semantic codec with $\{N_1,N_2,N_3,N_4\}=\{2,2,2,2\}$ Transformer blocks. For model training, we use variable learning rate, which decreases step-by-step from 1e-4 to 2e-5. The batchsize is set as 16. $\gamma$ is set as 1.15. The whole framework is optimized with Adam \cite{Adam} algorithm. All the experiments of MHPSC and other DL-based benchmarks are conducted in RTX3090 GPUs with Pytorch2.0.0.

\subsubsection{Comparison Benchmarks}
In the experiments, several benchmarks are given as below

$\textbf{ViTSC}$: The DL-empowered semantic communication framework \cite{mhdeepsc} in multi-hop wireless image transmission scenarios. Recursive training method is adapted for alleviating distortion accumulation.

$\textbf{WITT}$: The wireless image transmission transformer \cite{WITT} with Swin Transformer \cite{swin} as network backbone, along with the ChannelModnet for SNR-adaptive transmission.

$\textbf{ADJSCC}$: The adaptive deep JSCC scheme in \cite{ADJSCC} with the convolution neural network (CNN) structure, directly blending SNR as side information with original features in attention modules.

$\textbf{BPG+LDPC+QAM}$: The traditional coding transmission scheme with the Better Portable Graphics (BPG) \cite{bpg} coder as the source coding and the LDPC coder as the channel coding scheme, enhanced by the MMSE equalization for combating the Rayleigh fading channels. QAM is utilized as the modulation scheme.

Note that all the DL-based schemes employ the recursive training method. BPG coder is utilized through \cite{bpg} while 5G LDPC and QAM are deployed aided by sionna \cite{sionna}.  

\subsubsection{Evaluation Metrics}

We leverage the widely used pixel-wise metric peak signal-to-noise ratio (PSNR), perceptual-level multi-scale structural similarity (MS-SSIM) and learned perceptual image patch similarity (LPIPS) as measurements for the reconstructed image quality.  

\subsection{Results Analysis}

\subsubsection{Performance for Different SNRs}
We first evaluate the anti-noise performances of MHPSC under Rayleigh fading channels with a specific CBR, where $CBR=\frac{L+L_r}{H\times W\times 3} $. We set the whole CBR as 0.077 while the fixed CBR for the common semantic link is 0.062. As shown in Fig. \ref{fig_3}(a), it is clearly to observe that MHPSC outperforms all other benchmarks. Compared to other DL-based schemes, MHPSC outperforms WITT for over 1 dB in terms of PSNR on average, where the gap increases as the SNR value decreases. This trend verifies that parallel link-based MHPSC is more robust to the channel interference than single link-based schemes. With the parallel link, only limited increase of bandwidth cost compared to the original smenatic link enables such performance gain derived from distortion accumulation mitigation. For traditional separated coding schemes, 2/3 code rate LDPC with 16QAM and 1/2 code rate LDPC with 16QAM are set, which reflect corresponding anti-noise performances under different SNR levels. Compared to traditional schemes, MHPSC provides much more performance gain and stability since traditional schemes would be confronted with serious cliff effect with harsh channel conditions. As shown in Fig. \ref{fig_3}(b) and Fig. \ref{fig_3}(c), the DL-based schemes achieve better reconstruction results in terms of MS-SSIM and LPIPS compared to traditional SSCC schemes, which means that the satisfying visual perception quality is ensured through extracting semantics inside raw images. Compared to other schemes, MHPSC preserves even much more high frequency image details. These results demonstrate the effectiveness of MHPSC for multi-hop transmission under Rayleigh fading channels with various noise intensities.

\subsubsection{Performance for Different CBRs}
Then we evaluate the bandwidth compression performances of MHPSC under Rayleigh fading channels with SNR = 10 dB. As shown in Fig. \ref{fig_4}(a), MHPSC achieves much performance gain compared to other schemes. Compared to WITT, the performance gap increases as the decrease of CBRs. Such insight mainly lies intrinsic distortion accumulation problem in multi-hop scenarios. Transmitting semantics with small CBR results in low quality for image reconstruction. However, residual compensation mitigates the distortion at the end of each hop, greatly alleviating the whole distortion for multiple hops. For the visual perceptual-level indexes in Fig. \ref{fig_4}(b) and Fig. \ref{fig_4}(c), MHPSC retains relatively satisfying visual quality for a wide range of CBRs compared to other schemes. 
\begin{figure}[htbp]
	\centering
	\includegraphics[width=3.4in]{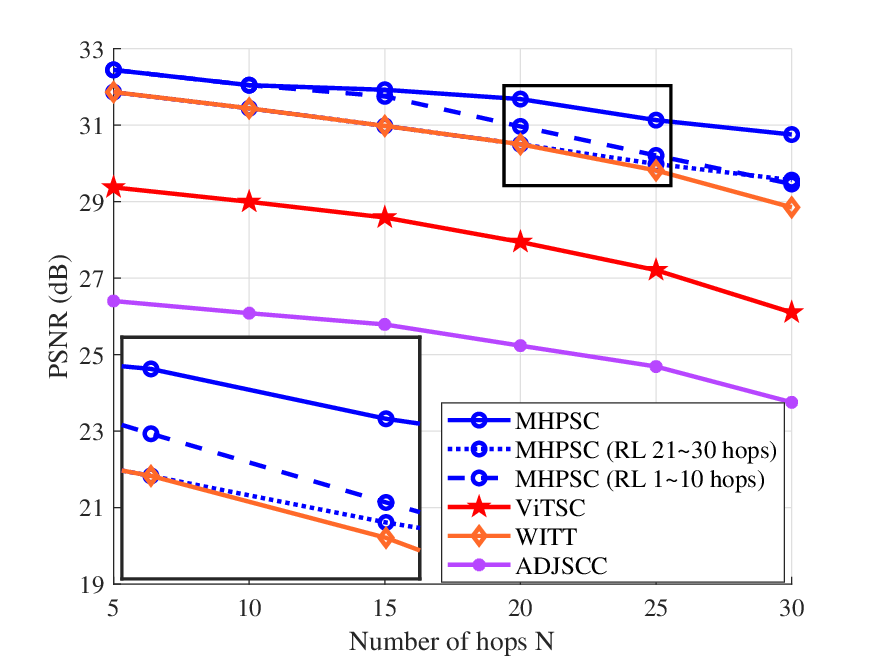}
	\caption{Performance of different hop numbers.}
	\label{fig_5}
\end{figure}

\subsubsection{Performances of different numbers of hops}
Finally, we evaluate the multi-hop transmission performance with different hop numbers. As shown in Fig. \ref{fig_5}, hop numbers are ranged from 5 to 30. It is clearly to observe that MHPSC surpasses all other DL-based schemes in terms of PSNR for various hop numbers. For proposed MHPSC, less performance degradation is presented with much large hop number such as $N=30$. While for other schemes, although the same recursive training method is adapted to promote each hop aware of previous hops, the reconstructed image quality degrades seriously with more hops. We further evaluate MHPSC by deploying its residual compensation link over different hop sequences. Specifically, we test two configurations. RL 1$\sim$10 refers to the application of parallel link only to the initial 10 hops, while RL 21$\sim$30 for only the final 10 hops. The results indicate that compensating the initial hops maintains stable performance for approximately 15 hops, after which it degrades as the hop count increases. In contrast, compensating the final hops gradually mitigates distortion accumulation, with performance improving as the hop count rises. Ultimately, by the 30-th hop, this late-stage compensation performs as effectively as the initial-stage configuration. With the above analysis, MHPSC is able to perform robust wireless image transmission in multi-hop scenarios.


\begin{thebibliography}{00}
	\bibitem{ADJSCC}
	J. Xu, B. Ai, W. Chen, et al. ``Wireless image transmission using deep source channel coding with 	attention modules", \emph{IEEE Trans. Circuits Syst. Video Technol.}, vol. 32, no. 4, pp. 2315-2328, Apr. 2022.

	\bibitem{LCFSC}
	B.~Xie, Y.~Wu, Y.~Shi, W.~Z, S.~Cui, and M.~Debbah, ``Robust image semantic coding with learnable CSI fusion masking over MIMO fading channels", \emph{IEEE Trans. Wireless Commun.}, vol. 23, no. 10, pp. 14155-14170, Oct. 2024.

	\bibitem{parasc}
	M. Xu et al., ``Semantic-aided Parallel Image Transmission Compatible with Practical System," \emph{IEEE Trans. Wireless Commun.} (early access), May 2025.

	\bibitem{relay}
	W. An, Z. Bao, H. Liang, C. Dong and X. Xu, ``A Relay System for Semantic Image Transmission Based on Shared Feature Extraction and Hyperprior Entropy Compression", \emph{IEEE Int. Things J.}, vol. 11, no. 9, pp. 16158-16170, May, 2024.

	\bibitem{mhdeepsc}
	G. Zhang, Q. Hu, Y. Cai and G. Yu, ``Alleviating Distortion Accumulation in Multi-Hop Semantic Communication", \emph{IEEE Commun. Lett.}, vol. 28, no. 2, pp. 308-312, Feb. 2024.

	\bibitem{hsc}
	H. Nam, J. Park, J. Choi, et al. ``Hybrid Semantic-Complementary Transmission for High-Fidelity Image Reconstruction", \emph{arxiv:2507.17196}, Jul. 2025. [Online]. Available: \url{https://arxiv.org/abs/2507.17196}.

	\bibitem{WITT}
	K.~Yang, S.~Wang, J.~Dai, et al. ``WITT: A wireless image transmission transformer for semantic communications", \emph{IEEE Int. Conf. Acoust. Speech Signal Process. (ICASSP)}, Rhodes Island, Greece, Jun. 2023, pp. 1-5.
	
	\bibitem{rc}
	F.~ Mentzer, L.~ Gool, M.~ Tschannen, ``Learning better lossless compression using lossy compression", in \emph{IEEE Conf. Comput. Vis. Pattern Recognit. (CVPR)}, online, Oct. 2020, pp. 6638-6647.

	\bibitem{WVSC}
	B. Xie, Y. Wu, Y. Shi, B. Feng, W. Zhang, J. Park, and T. Quek, ``WVSC: Wireless Video Semantic Communication with Multi-frame Compensation", \emph{arxiv:2503.21197}, Mar. 2025. [Online]. Available: \url{https://arxiv.org/abs/2503.21197}.

	\bibitem{cvst}
	B. Xie et al., ``Context Video Semantic Transmission with Variable Length and Rate Coding over MIMO Channels," Dec. 2025. [Online]. Available: \url{https://arxiv.org/abs/2601.06059}.

	\bibitem{UDIS-D}
	L.~Nie, C.~Lin, K.~Liao, et al. ``Unsupervised deep image stitching: Reconstructing stitched features to images", \emph{IEEE Trans. Image Process.}, vol. 30, pp. 6184--6197, Jul. 2021.

	\bibitem{swin}
	Z. Liu, Y. Lin, Y. Cao, H. Hu, Y. Wei, Z. Zhang, S. Lin, and B. Guo, ``Swin transformer: Hierarchical vision transformer using shifted windows", in \emph{IEEE Conf. Comput. Vis. Pattern Recognit. (CVPR)}, Montreal, QC, Canada, Oct. 2021, pp. 9992-10002.
	
	\bibitem{Adam}
	D. ~Kingma and J.~Ba, ``Adam: A method for stochastic optimization", \emph{arxiv:1412.6980}, Dec. 2014. [Online]. Available: \url{https://arxiv.org/abs/1412.6980}.

	\bibitem{bpg}
	F.~Bellard, ``BPG Image Format.", Accessed: Apr. 2018. [Online]. Available: \url{https://bellard.org/bpg/}.

	\bibitem{sionna} 
	H., Jakob, et al. ``Sionna: An open-source library for next-generation physical layer research", Mar. 2022. [Online]. Available: \url{https://arxiv.org/abs/2203.11854}.
\end{thebibliography}
\end{document}